# Origin of high piezoelectric response of Pb(Zr$_x$Ti$_{1-x}$)O$_3$ at the morphotropic phase boundary: Role of elastic instability

Akhilesh Kumar Singh, Sunil Kumar Mishra, Ragini and Dhananjai Pandey[a)]

School of Materials Science & Technology, Institute of Technology,

Banaras Hindu University, Varanasi-221 005, INDIA

Songhak Yoon and Sunggi Baik

Department of Materials Science and Engineering,

Pohang University of Science and Technology, Pohang 790-784, Korea

Namsoo Shin

Pohang Accelerator Laboratory, Pohang University of Science and Technology,

Pohang 790-784, Korea

**Abstract**

Temperature dependent structural changes in a nearly pure monoclinic phase composition (x=0.525) of Pb(Zr$_x$Ti$_{1-x}$)O$_3$ (PZT) have been investigated using Rietveld analysis of high-resolution synchrotron powder x-ray diffraction data and correlated with changes in the dielectric constant and planar electromechanical coupling coefficient. Our results show that the intrinsic piezoelectric response of the tetragonal phase of PZT is higher than that of the monoclinic phase. It is also shown that the high piezoelectric response of PZT may be linked with an anomalous softening of the elastic modulus $(1/S_{11}^E)$ of the tetragonal compositions closest to the morphotropic phase boundary.

[a)] Author to whom correspondence should be addressed;
electronic mail: dpandey@bhu.ac.in

A unique feature of the phase diagrams of the technologically important solid solution systems between the ferroelectric $PbTiO_3$ and the antiferroelectric $PbZrO_3$ or relaxor ferroelectrics like $Pb(Mg_{1/3}Nb_{2/3})O_3$ and $Pb(Zn_{1/3}Nb_{2/3})O_3$ is the existence of a nearly vertical morphotropic phase boundary (MPB) for which the dielectric and piezoelectric properties show exceptionally high values [1]. There is no satisfactory explanation for this phenomenon. The currently most acceptable model of high-piezoelectricity in these materials is based on the premise that the recently discovered intermediate monoclinic phases [2-7] can mediate the rotation of the polarization vector from the [111] polarization direction of the rhombohedral phase to the [001] of the tetragonal phase [8-9]. The results of a high-resolution x-ray powder diffraction study on sintered $Pb(Zr_xTi_{1-x})O_3$ PZT samples [10] shows that the field induced piezoelectric elongations in the tetragonal and rhombohedral compositions occur not along the polar [001] and [111] directions but along [101] and [001] directions, respectively, corresponding to the atomic displacements in the monoclinic Cm phase, have been taken as evidence for the polarization rotation model. The first principle calculations by Bellaiche et al [11] suggest that the monoclinic phase possesses intrinsically higher piezoelectric properties, as compared to the tetragonal and rhombohedral phases of neighbouring compositions. However, no attempt has so far been made to compare the piezoelectric properties of the monoclinic phase of PZT with those of the tetragonal and rhombohedral phases for the same composition. In the present work, we have addressed this question experimentally for the tetragonal and monoclinic phases of the PZT system. Much to our surprise, we find that it is the tetragonal phase, and not the monoclinic phase, which has got superior piezoelectric properties. The higher piezoelectric response of PZT near MPB is found to be linked with an anomalous elastic softening around the room temperature for the tetragonal compositions close to the MPB.

Chemically homogeneous and highly stoichiometric PZT powders were prepared by the semi-wet route developed by Singh et al [12]. High-resolution synchrotron x-ray diffraction (XRD) data was recorded using 8C2 HRPD beamline at Pohang Light Source (PLS), Pohang Accelerator Laboratory, Pohang, Korea. For dielectric and piezoelectric characterizations, sintered PZT pellets with densities greater than 98% were electroded with fired-on silver paste. Poling was carried out by applying 20 kV/cm dc field for 50 minutes at 373K in silicon oil bath. Dielectric measurements on unpoled and poled samples were carried out at 100 KHz using a Schlumberger (SI 1260) impedance/gain phase analyzer. The planar electromechanical coupling coefficient ($k_P$) was determined by following the procedure described in Ref.1. FULLPROF [13] program was used for Rietveld analysis of the XRD data. Anisotropic peak broadening functions due to Stephens [14] were used to model peak profiles. Background was modeled using fifth order polynomial. In agreement with the earlier reports [2-3], anisotropic thermal parameters were found necessary for the Pb atom in the refinements; for the remaining atoms isotropic thermal parameters were found adequate.

We have recently shown that the structure of PZT is tetragonal (space group P4mm), and monoclinic (space group Cm) for $x \leq 0.515$ and $x \geq 0.525$ while the two phases coexist for x=0.520 [15]. The monoclinic structure of x=0.525 transforms to the tetragonal structure on heating above room temperature due to the tilted nature of the MPB towards the Zr-rich region [1]. Fig.1 depicts the XRD profiles of the pseudocubic 110, 111 and 200 reflections along with the Rietveld fits obtained from full pattern refinements in the two-theta range 18 to 130 degrees at three representative temperatures. The observed and calculated profiles match extremely well using monoclinic (Cm) and tetragonal (P4mm) phase models at 300K and 570 K (see Fig. 1(a) and (c)). For the temperature range (350K to 550 K) across the monoclinic to tetragonal phase transition temperature, the monoclinic and the tetragonal phases coexist, as can be seen from the excellent Rietveld fits to the observed XRD data for a representative temperature (400K) in Fig.1(b). The evolution of the unit cell parameters as a function of temperature, as obtained by Rietveld analysis of the XRD data at each temperature, is shown in Fig. 2 which reveals monoclinic to tetragonal phase transition around 490K and tetragonal to cubic transition around 650K. Both the transitions are found to be first order, as inferred from the coexistence of the low and high temperature phases across the phase transition temperature.

The monoclinic to tetragonal phase transition is accompanied with an anomaly in the dielectric constant which is found to occur at the same transition temperature (~493K) for both unpoled and the poled samples (see Fig.3(a) and (b) for x=0.525). Thus poling does not affect the monoclinic to tetragonal phase transition temperature. The monoclinic to tetragonal phase transition temperature obtained by dielectric measurements is in close agreement with the transition temperature (~490K) obtained from the XRD studies (see Fig.2). The temperature variation of the planar electromechanical coupling coefficient ($k_P$) and piezoelectric strain coefficient ($d_{31}$) for x=0.525 is shown in Fig. 4(a). The $d_{31}$ values were obtained using Eq.6 on page 293 of Ref.1. It is evident form this figure that on heating above the room temperature (~300K), the values of $k_P$ and $d_{31}$ gradually increase and peak at the monoclinic to tetragonal phase transition temperature of ~490K.

Above the transition temperature, $k_P$ and $d_{31}$ finally stabilize to a nearly temperature independent value of ~0.44 and ~125 pC/N, respectively, in the tetragonal phase. These values are significantly higher than the room temperature values of $k_P$~0.35 and $d_{31}$~55 pC/N for the monoclinic phase. This shows that the piezoelectric response of the tetragonal phase is distinctly higher than that of the monoclinic phase. The intrinsic difference between the $k_P$ and $d_{31}$ values of the tetragonal and monoclinic phases may be higher than ~0.1 and 70, respectively, since at high temperatures (above the tetragonal to monoclinic transition temperature) partial depoling can lead to the deterioration of the piezoelectric response of the tetragonal phase. Fig.4(a)  shows one more peak in $k_P$ versus temperature plot below room temperature which is linked with the phase transition from the monoclinic Cm phase to a superlattice monoclinic phase with Cc space group [4-5]. We find that higher piezoelectric response of the tetragonal phase vis-à-vis the monoclinic phase is a common feature of other tetragonal compositions as well,  as can be seen from Fig.4(b) and (c) for x=0.520 and 0.515. The $k_P$ for all the three compositions shown in Fig.4 is nearly temperature independent above the monoclinic Cm to tetragonal phase transition temperature and its value is significantly higher than that of the Cm phase.

The clue to this higher piezoelectric response of the tetragonal phase, as compared to the monoclinic Cm phase, lies in the softening of some elastic modulii on approaching the tetragonal to monoclinic phase transition temperature. The elastic modulus ($1/S_{11}^{E}$) of the poled PZT ceramics can be estimated from the piezoelectric resonance frequency ($f_r$) using the relationship, $1/S_{11}^{E} = \pi^2 d^2 f_r^2 (1-\sigma^{E2})\rho/\eta_1^2$, where d is the diameter of the pellet, $\rho$ is the density and $\sigma^{E}$ (=0.31) and $\eta_1$ (=2.05) are constants [1]. The variation of $1/S_{11}^{E}$ with temperature is shown in Fig.5 for x=0.515, 0.520, 0.525 and also for a pseudo-rhombohedral composition x=0.550 (whose correct space group is Cm (Ref.15)), which transforms to the cubic phase without any intermediate tetragonal phase. It is evident from Fig.5(c) and (d) that the $1/S_{11}^E$ of the tetragonal phase decreases with decreasing temperature upto the tetragonal to monoclinic (Cm) phase transition temperature, as expected for a soft mode system. After the transformation to the monoclinic phase, the elastic modulus hardens in the Cm phase region until another instability corresponding to the Cm to the superlattice Cc phase transition sets in when the elastic modulus again starts decreasing with decreasing temperature. In the Cc phase, the normal hardening of the elastic modulus is restored. For the composition x=0.525 also, the $1/S_{11}^E$ above the tetragonal to monoclinic phase transition temperature of ~490K is anomalous, as it does not increase with decreasing temperature. Since the tetragonal-monoclinic phase transition of PZT is closest to room temperature for x=0.520, the elastic softening effects shall also be most pronounced for this composition at room temperature. We believe that the highest piezoelectric response of PZT for x=0.520 is linked with this anomalous elastic softening. A small electric field applied to such an elastically soft solid can produce large piezoelectric strain through the electromechanical coupling. In contrast to the tetragonal compositions at room temperature, the monoclinic and pseudo-rhombohedral compositions do not exhibit elastic softening in the vicinity of the room temperature (see Fig. 5(a) and (b) for x=0.550 and 0.525) as a result of which their piezoelectric properties are inferior to those of the tetragonal compositions near the MPB. In addition to the piezoelectric properties, the composition (x) dependence of the dielectric constant also shows a peak at x ≈ 0.52, which may be linked with the softening of the zone centre E(TO) mode frequency ω(x) on approaching the MPB from the tetragonal side, as has been pointed out earlier on the basis of Raman scattering studies [16].

To summarize, we have correlated the structural changes associated with the monoclinic Cm to tetragonal phase transition in PZT with the change in the electromechanical response. We have presented unambiguous evidence for higher electromechanical response of the tetragonal phase than that of the monoclinic which contradicts the predictions of the first principle calculations about higher piezoelectric response of the monoclinic phase [8, 11]. The high piezoelectric response of PZT at the MPB is shown to be linked with the elastic instability of the tetragonal compositions and the proximity of the tetragonal to monoclinic phase transition temperature to room temperature for x=0.520.

**Acknowledgements:** The work at Pohang Accelerator Laboratory has been supported financially by MOST and Brain Korea 21 Program. The experiments at Pohang Light Source (PLS) were also supported by the Ministry of Science and Technology (MOST) and POSTECH, Pohang, Korea.

**Figure Caption:**

**Fig.1** Temperature evolution of the 110,111 and 200 pseudocubic powder synchrotron x-ray diffraction profiles of $Pb(Zr_{0.525}Ti_{0.475})O_3$. The solid dots show the observed diffraction profiles, while the continuous line the calculated patterns obtained by the Rietveld analysis of the data for different structures. The vertical tick marks show the positions of various Bragg reflections.

**Fig.2** Evolution of the lattice parameters with temperature obtained after Rietveld analysis of the synchrotron XRD data for $Pb(Zr_{0.525}Ti_{0.475})O_3$. For easy comparison, the equivalent perovskite cell parameters $a_m$ and $b_m$ calculated from the monoclinic cell parameters $A_m$ and $B_m$, are plotted in the monoclinic region ($a_m = A_m/\sqrt{2}$, $b_m = B_m/\sqrt{2}$).

**Fig. 3** Temperature variation of the real part of the dielectric constant ($\varepsilon'$) for $Pb(Zr_xTi_{1-x})O_3$ ceramics with x=0.525 (a) Poled and (b) unpoled sample.

**Fig.4** Temperature variation of the planar electromechanical coupling coefficient ($k_P$) of $Pb(Zr_xTi_{1-x})O_3$ ceramics: (a) x=0.525 (b) x=0.520 (c) x=0.515. For x=0.525, the temperature dependence of $d_{31}$ is also shown in (a).

**Fig.5.** Temperature variation of the elastic modulus ($1/S_{11}^E$) of $Pb(Zr_xTi_{1-x})O_3$ ceramics (a) x=0.550 (b) x=0.525 (c) x=0.520 (d) x=0.515.

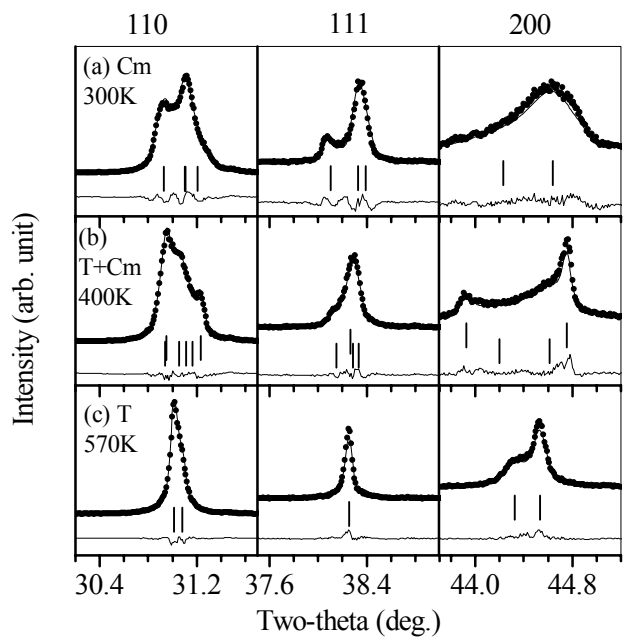

110                111                200

(a) Cm
300K

(b)
T+Cm
400K

(c) T
570K

Intensity (arb. unit)

30.4        31.2    37.6    38.4    44.0    44.8

Two-theta (deg.)

Fig.1

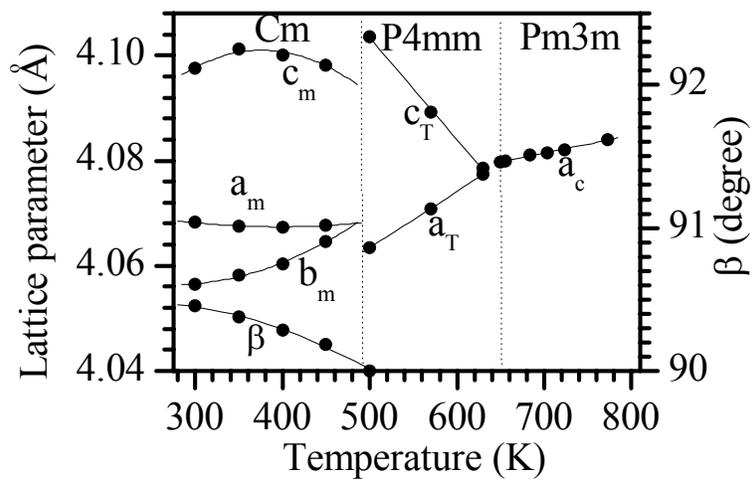

Fig.2

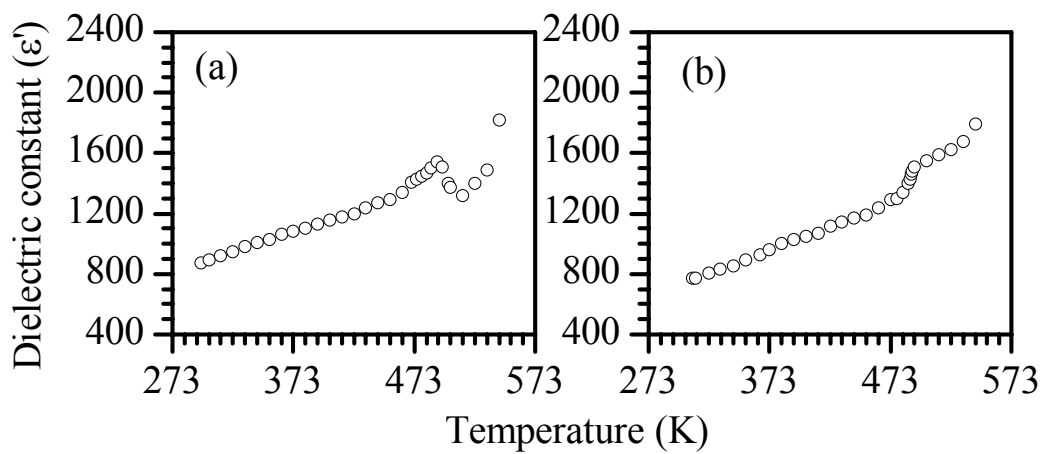

Fig.3

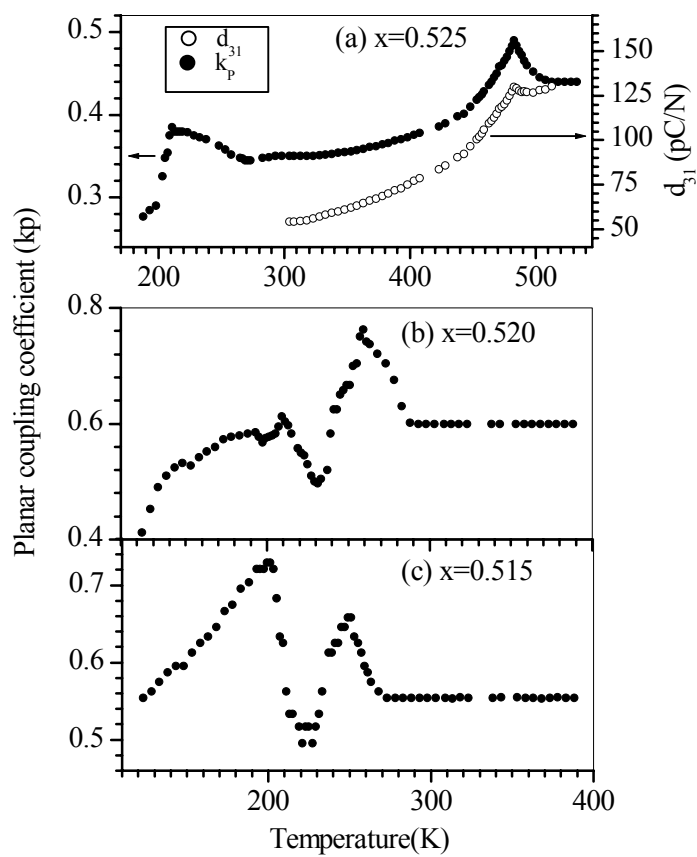

Fig.4

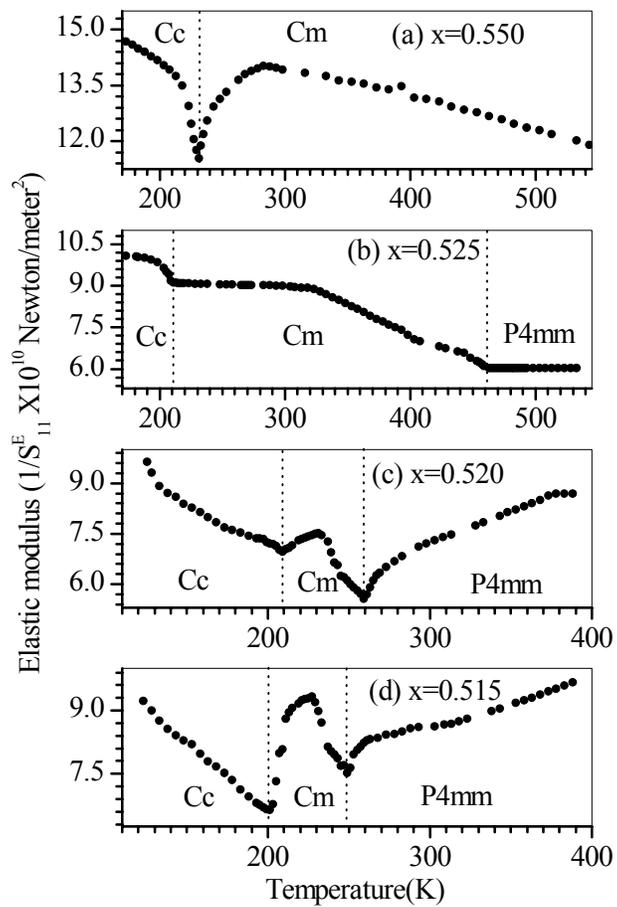

Fig.5